\documentclass{article}
\UseRawInputEncoding
\usepackage{CPNS_Paper}
\usepackage{graphicx,url}
\usepackage[utf8]{inputenc}
\usepackage{amssymb,latexsym,amsmath,amsthm}
\usepackage{graphics}
\usepackage{amsfonts}
\usepackage{float}
\usepackage{color}
\usepackage{graphicx}
\usepackage{indentfirst}
\usepackage{dcolumn}
\usepackage{bm}
\usepackage{newtxtext, newtxmath}
\DeclareGraphicsExtensions{.pdf,.jpeg,.ps, .png, .eps, .tiff}
\usepackage{epstopdf}
\usepackage{lipsum}
\usepackage{tabularx}
\usepackage{booktabs}
\usepackage{array}
\usepackage{xcolor}
\usepackage{grffile}
\usepackage{boxhandler}
\usepackage{afterpage}
\usepackage{url}
\usepackage{subfig}

\title{Contagion-Preserving Network Sparsifiers: Exploring Epidemic Edge Importance Utilizing Effective Resistance}

\author{Alexander Mercier\inst{1,2}}

\address{
  University of South Florida
\nextinstitute
  Santa Fe Institute
}

\begin{document} 

\maketitle

\begin{abstract}

Network epidemiology has become a vital tool in understanding the effects of high-degree vertices, geographic and demographic communities, and other inhomogeneities in social structure on the spread of disease. However, many networks derived from modern datasets are quite dense, such as mobility networks where each location has links to a large number of potential destinations. One way to reduce the computational effort of simulating epidemics on these networks is sparsification, where we select a representative subset of edges based on some measure of their importance. Recently an approach was proposed using an algorithm based on the effective resistance of the edges. We explore how effective resistance is correlated with the probability that an edge transmits disease in the SI model. We find that in some cases these two notions of edge importance are well correlated, making effective resistance a computationally efficient proxy for the importance of an edge to epidemic spread. In other cases, the correlation is weaker, and we discuss situations in which effective resistance is not a good proxy for epidemic importance.
\end{abstract}

\section{Introduction}

\subsection{Motivation}

Networks arise in a variety of contexts, from the study of epidemics, social contagions, terrorism, and biological invasions, to how knowledge itself is organized through epistemological networks. Use of network-based models for simulating epidemics has become particularly popular. However, simulating a stochastic epidemic model on a large network is computationally expensive, especially for dense networks, such as those derived from high-resolution mobility data which have been increasingly used in modeling contagion spread \cite{Halloran:1, Oliver:1, Wesolowski:1}. In these networks, there is a link between every pair of destinations, with weights corresponding to the flow of people who travel between them \cite{Wesolowski:1}. Considering all possible links as potential paths of infection takes significant of computation, and the problem is exacerbated by the need to perform many independent runs to get a sense of the probability distribution of epidemic sizes in stochastic models, as well as to test the effect of various intervention strategies. It is common to apply naive heuristics, like simply removing links whose weights are below some threshold, but it is not clear to what extent this preserves the true behavior of an epidemic, since rare events on low-weight edges can have important downstream consequences. 

\subsection{Sparsification}
One way to address this is \emph{sparsification}: choosing a subset of important links in the network, deriving a sparser network whose behavior would be faithful to the original but which is far less costly to study. The aim of sparsification is to approximate a network, $G(V,E,\phi)$ by a graph sparsifier, $\Tilde{G}(V,\Tilde{E},\Tilde{\phi})$, on the same set of vertices, $V$, but with a reduced number of edges, $\Tilde{E}$, and modified edge weights, $\Tilde{\phi}$  such that $\Tilde{G}$ approximates G in some appropriate metric or metrics. Since graphs arise in the study of complex networks, graph sparsification has become both a topically important area of study and an interesting mathematical challenge. 

Therefore, network sparsification for networks used in stochastic epidemic simulations are motivated by two primary aspirations. First, to lower the computational cost of simulating epidemics on the network while retaining the same average dynamics. Second, the underlying goal is for the sparsification algorithm to conserve edges important to epidemic spread and remove edges that are not. Likewise, when paired with associated metadata, the removal of “unimportant” edges and the conservation of “important” edges may allow further analytic insight into network topological structure. We explore the notion of a contagion-preserving network sparsifier (CPNS) which seeks to reduce the number of edges in a network while simultaneously approximating average epidemic dynamics. In this way, a CPNS reduces the computational costs incurred in dynamical simulations on $G$, by approximating typical dynamics on $\Tilde{G}$. Yet, how do we determine which links of the original network are the most important in an epidemic? 

One possible approach comes from an algorithm created by Daniel Spielman and Shang-Hua Teng, for which they won the G\"{o}del Prize in 2015, and a simplification and improvement by Spielman and Srivastava \cite{Spielman:1, Spielman:2}. The idea from the Spielman--Srivastava algorithm is to randomly sample edges with probability proportional to their effective resistance: in physical terms, the potential difference between their endpoints when one unit of current flows between them (and where each edge has resistance equal to the reciprocal of its weight). Effective resistance, also called "current-flow betweenness" or "spanning edge betweenness," has been explored as a measure of edge importance \cite{Teixeira:1,Teixeira:2,Bozzo:1}. This resistance is high if an edge is one of the only ways to quickly get from one part of the network to another or if alternate paths are long or consist of low-weight edges. Choosing edges this way preserves important aspects of the graph spectrum --- approximately preserving the graph Laplacian --- and makes it possible to solve certain systems of equations in nearly-linear time \cite{Spielman:2}. However, this notion of “importance” may or may not align with the role edges play in an epidemic, since approximately preserving the graph Laplacian (which governs linear dynamics on the network) might not preserve the highly nonlinear dynamics of an epidemic. Are the edges with higher probability of selection in the Spielman--Srivastava algorithm more likely to spread disease? What is the right way to sparsify a network if our goal is to preserve its epidemic behavior rather than its spectrum?

\subsection{Methodology}

Towards this end, we develop the novel concept of an infection spanning tree, from which a general notion of epidemic edge importance may be formed and contrasted against effective resistance. Another formulation of effective resistance is the probability that a random spanning tree includes a given edge, where the spanning tree is chosen uniformly or with probability proportional to the product of edge weights. In contrast, the infection spanning tree lets us measures which trees and paths are most likely to spread a contagion.

\newpage

Focusing on SI contagion dynamics with a discrete-time SI model, we use the Spielman--Srivastava algorithm utilizing effective resistance in $O(n\log^cn)$ time to produce CPNS, drawing parallels between the linear flow conceptualization of a network and contagion spread on that network.  This research takes four approaches not taken in previous literature \cite{Swarup:1}. Swarup et al. also used effective resistance. However, in comparing with epidemics on the original network they used the metric of minimum Hamming distance \cite{Swarup:1}. We use a suite of metrics to study the performance of CPNS, namely \emph{average} Hamming distance, mutual information, and how well the fraction of infected vertices over time matches the epidemic on the original network. Second, we compare effective resistance with the epidemic edge importance using infection spanning trees. Third, while preceding work centered  around aggregate SI dynamics on CPNS, this research examines SI contagion dynamics on CPNS primarily through time. Lastly, we compare the Spielman--Srivastava algorithm with a simpler method that samples edges uniformly. 

In order to explore important edges in contagion spread and CPNS, we conduct a range of experiments on four random networks and a real-world air transport network. The results will show that while the linear flow analog to contagion processes is conceptually fertile, a range of diverse metrics must be implemented in order to view the full picture of the effectiveness of a given CPNS. This spectral sparsification algorithm can be used to create effective CPNS, permitting the removal of $75\%$ of edges in some networks while approximately preserving the same average SI dynamics through time. However, more research must be conducted to understand the importance of any given edge within the context of an epidemic in order to fully grasp the workings of a CPNS.

\section{Methods} \label{sec:firstpage}

\subsection{Linear Flow and Contagion Spread}

\begin{table}[h!]
  \begin{center}
    \begin{tabular}{|l|c|} % <-- Alignments: 1st column left, 2nd middle and 3rd right, with vertical lines in between
      \hline
      \textbf{Current Flow} & \textbf{Contagion Spread}\\
      \hline
      \text{Conductance} & \text{Probability of Transmission Along An Edge}\\
      \hline
      \text{Resistance} & \text{Expected Time to Infection}\\
      \hline
      \text{Current} & \text{Expected Contagion Flow}\\
      \hline
    \end{tabular}
  \end{center}
  \caption{Linear Flow and Contagion Processes}
\end{table}

We begin by noting the parallels between linear flow in electrical networks and epidemic processes on social contact networks (Table 1). In contagion processes, we assume that transmission can occur along each edge independently. Therefore, the probability of the contagion spreading along an edge, $\pi_e$ is treated as analogous to the conductance of that edge. The reciprocal of conductance, known as resistance, is the expected time to infection along that edge or commute time, $T_e$. The flow of a contagion along the network can then be thought of as current flow on the corresponding electrical network. Effective resistance, $R_e$, is then defined as the potential difference across an electrical network when one unit of current is injected at a vertex, $i$, and extracted at another vertex, $j$, taking into account all possible paths between $i$ and $j$. In order to approximate the effective resistance for all pairs of vertices, we created an implementation in R of the Spielman--Srivastava algorithm. Specifically, we implemented the formulation by Koutis, Levin, and Peng which works in time nearly linear to the number of edges \cite{Koutis:1}.

Effective resistance between any two given vertices is given by the graph Laplacian when the resistance of each edge is defined as the inverse of its weight. The effective resistance between $i$ and $j$ is given by

\[ R_{ij} = (e_i - e_j)^T L^+ (e_i-e_j) \]

where $L^+$ is the \textit{pseudoinverse} of the graph Laplacian and $e_i$ is the column vector where there is a $1$ at $i$ and zero elsewhere. The algorithm to approximate effective resistance between all vertex pairs inverts the Laplacian approximately using a random projection technique based upon the Johnson-Lindenstrauss lemma \cite{Spielman:2}. This approximation guarantees the approximate effective resistance given by the algorithm is between $(1-\varepsilon)R_{ij}$ and $(1+\varepsilon)R_{ij}$ for some constant error parameter $\varepsilon$ which can be made as small as desired. In our implementation we typically have $\varepsilon \le 0.1$.

\subsection{Sparsification using Effective Resistance}

\noindent\rule{15cm}{0.6pt}

\noindent\textbf{Algorithm 1: Sparsification by effective resistance \cite{Spielman:1}}

\noindent\rule{15cm}{0.6pt}

\noindent\textbf{Input:} network $G(V,E,\phi)$\\
\textbf{Output:} network $\Tilde{G}(V,\Tilde{E},\Tilde{\phi})$\\
\textbf{Parameters:} $q$, the number of samples 

\noindent\textbf{Procedure:}\\
\noindent Choose random edge $e$ from $G$ with probability $p_{e}\propto w_eR_{\text{e}}$\\
Add edge $e$ to $\Tilde{G}$ with weight $\Tilde{w}_{e} = w_{e}/p_{e}q$\\
Take $q$ samples with replacement and sum weights if an edge is chosen more than once

\noindent\rule{15cm}{0.6pt}

Sparsification via effective resistance approximately preserves the effective resistance between all vertices, ensuring the expected time to infection for any vertex remains approximately the same \cite{Swarup:1}. This is due to the fact the spectrum of the graph will remain similar to that of the original graph; the spectrum of a graph has been shown to govern many aspects of the diffusion on that graph \cite{Ganesh:1}. The Spielman--Srivastava Algorithm takes the effective resistance of all edges of a graph and uses $R_e$ to sample edges with probability, $p_e$, equal to $w_e R_e$ where $w_e$ is the weight of that edge. Notably, according to Algorithm 1, edge weights are sampled proportional to $\pi_e T_e$ \cite{Swarup:1}. An edge selected by the Spielman--Srivastava algorithm is assigned a new weight of $\Tilde{w}_e$ equal to $w_e/p_eq$ such that $q$ is the total number of samples taken. Edges are sampled with replacement, so that they can be selected multiple times. If the same edge is selected more than once, the edge weights are added together. Since the expected number of times $e$ is selected is $q p_e$, the expectation of $\Tilde{w}_e$ is equal to its original edge weight $w_e$. Thus, the expected adjacency matrix will equal the original adjacency matrix and the expected graph Laplacian will be equal to the original graph Laplacian. In this way, Algorithm 1 modifies edge weights to compensate for reduced edge number and can be thought of as being a part of a more general sampling strategy whereby edges are assigned probabilities by some metric of edge importance. 

Broadly, if there is an edge $e$ which connects vertices $i$ and $j$ such that no alternate paths exist between $i$ and $j$, or more generally if the alternate paths are long or involve lower-weight edges, then $T_e$ will be equal to the resistance of that edge with $\pi_eT_e$ equal to $1$. Conversely, if more paths are added between $i$ and $j$, then the expected time to infection decreases and $\pi_eT_e$ becomes less than 1. This suggests that as more paths are added between $i$ and $j$, there is less incentive for the Spielman--Srivastava algorithm to select $e$ for the sparsifier. This notion is similar to the concept of the "embeddedness" of an edge from Schaub et al. which is defined as $(1-\pi_eT_e)$ and conveys how important an edge is in weighted cuts of that graph or how much an edge acts as a "bottleneck" on that network \cite{SCHUAB:1}. 

\subsection{Contagion-Preserving Network Sparsifiers Methodology and Metrics}

In order to gauge the success of a CPNS, SI discrete-time process on the original network and CPNS are saved as an indexed list of strings. The string is of length $N$ where $N$ is the number of vertices within the network. All vertices are indexed $1$ through $N$ whereby entry $n$ of the string denotes vertex $n$ within the network and can be either $0$ or $1$ in the string, representing a susceptible or infected vertex, respectively. The cardinally of the indexed list of strings is $T$ where $T$ is the number of timesteps designated to run the model. The first string of the indexed list corresponds to the state of the SI model at timestep $1$, the second string corresponding to timestep $2$, and so on. The probability of an edge with weight $w_e$ transmitting a contagion with probability of transmission $\gamma$ is given by 
\[ \pi_e = 1-(1-\gamma)^{w_e} \]Because we are utilizing a discrete-time SI model, a contagion might be transmitted to a new vertex by two or more of its edges simultaneously, i.e. on the same time step. In this case, the edge that has the opportunity to transmit first is chosen uniformly at random. 

We employ the following set of metrics between the original network and CPNS contagion processes: average Hamming distance, mutual information score, and fraction of infected vertices in the network. The SI model is examined through time, where each metric is computed per time step. When contagion processes on the initial network are compared to those on the CPNS,  the same patient zero is selected. 

Because the SI model is stochastic, the purpose of the CPNS is not to precisely mimic the progress of any one run of the epidemic: even independent runs of the SI model on the original network will vary and have some typical, nonzero Hamming distance and mutual information. Then, it is appropriate for the CPNS to be evaluated on its preservation of average metrics over multiple runs. Therefore, we begin by calculating a baseline. This baseline is computed by averaging the respective metric over multiple runs on the original network. Additionally, we pick a small number of CPNS to generate and run multiple runs of the SI model and take the average of the Hamming distance, mutual information, and fraction of infected. To ensure that the CPNS is robust, we pick uniformly at random one patient zero to begin the simulation, keeping the same patient zero for both the runs on the original network and the CPNS. For each average metric, a $95\%$ confidence interval is also calculated. An effective CPNS will remain “close” to the given metric’s baseline while matching its confidence interval. 

\newpage

To better determine if an improved metric score produced by a CPNS is due to the addition of edges important to contagion spread or is merely due to the addition of more edges to the CPNS, we include a null model with which to compare. The null model uses the same framework as the Spielman--Srivastava algorithm, but samples uniformly from the set of edges with replacement. It will be entitled “uniform sampling.”  If the Spielman--Srivastava Algorithm is selecting edges of relevance to contagion spread, then the Spielman--Srivastava Algorithm CPNS should perform better than the corresponding uniform sampling CPNS while also selecting less edges. For easy comparison across multiple types of networks with a varying number of edges and to control edge number between the uniform sampling and SS sampling procedures, a CPNS with $25\%$, $50\%$ and $75\%$ of the total edges will be created for each network using both sampling procedures.

\subsection{Epidemic Edge Importance and Networks}

To investigate the relationship between effective resistance, the Spielman--Srivastava sparsification algorithm, and the spread of contagion, we introduce the measure of epidemic edge importance and an infection spanning tree. An infection spanning tree is created by simulating an SI contagion process from a patient zero until all vertices of the same component as patient zero are infected, keeping track of edges that transmit the contagion. Those edges which transmit the contagion make up the infection spanning tree. This process is performed over multiple runs, considering all possible patient zeros in the given network; the probability that an edge is found in any given infection spanning tree is termed the epidemic edge importance. The probability that an edge is selected by the Spielman--Srivastava algorithm, $w_eR_e$, and epidemic edge importance of all edges are normalized and organized in a Q-Q scatter plot, whereby the Pearson correlation coefficient can be used as a quantitative measure of similarity between the two metrics of edge importance. 

To explore the notion of edge importance as it relates to effective resistance, we test this methodology on four random networks and a real world airline network. The four random networks are a random network drawn from the configuration model with exponential logarithmic degree distribution, stochastic block model, complete network with edge weights drawn from a normal distribution, and a complete network with edge weights drawn from a power law distribution. The configuration network and stochastic block model each have $500$ vertices and both complete networks have $100$ vertices, denoted $K_{100}$. The configuration network and stochastic block model both have all edge weights set to one. The airline network (AirNet) contains $500$ vertices corresponding to the $500$ airports with the most traffic during the year 2002 in the United States\cite{Colizza:1}. Edge weights correspond to the total number of seats passing between a pair of airports \cite{Colizza:1}. 

\section{Results}
\subsection{Contagion-Preserving Network Sparsifiers}

We wish to check if the Spielman--Srivastava algorithm CPNS is close to the dynamics on the original network for average Hamming distance, mutual information, and fraction of vertices infected by the SI model as a function of time . Because the SI model is stochastic, the objective is not for the Hamming distance to be zero. Rather, it should be comparable to what is outputted from multiple independent runs on the original network. Likewise, for both mutual information and fraction of infected. For comparison, we also measure these quantities for CPNS with same percentage of total edges produced by uniform sampling. 

\begin{figure}[H]
    \centering
    \includegraphics[width=\columnwidth]{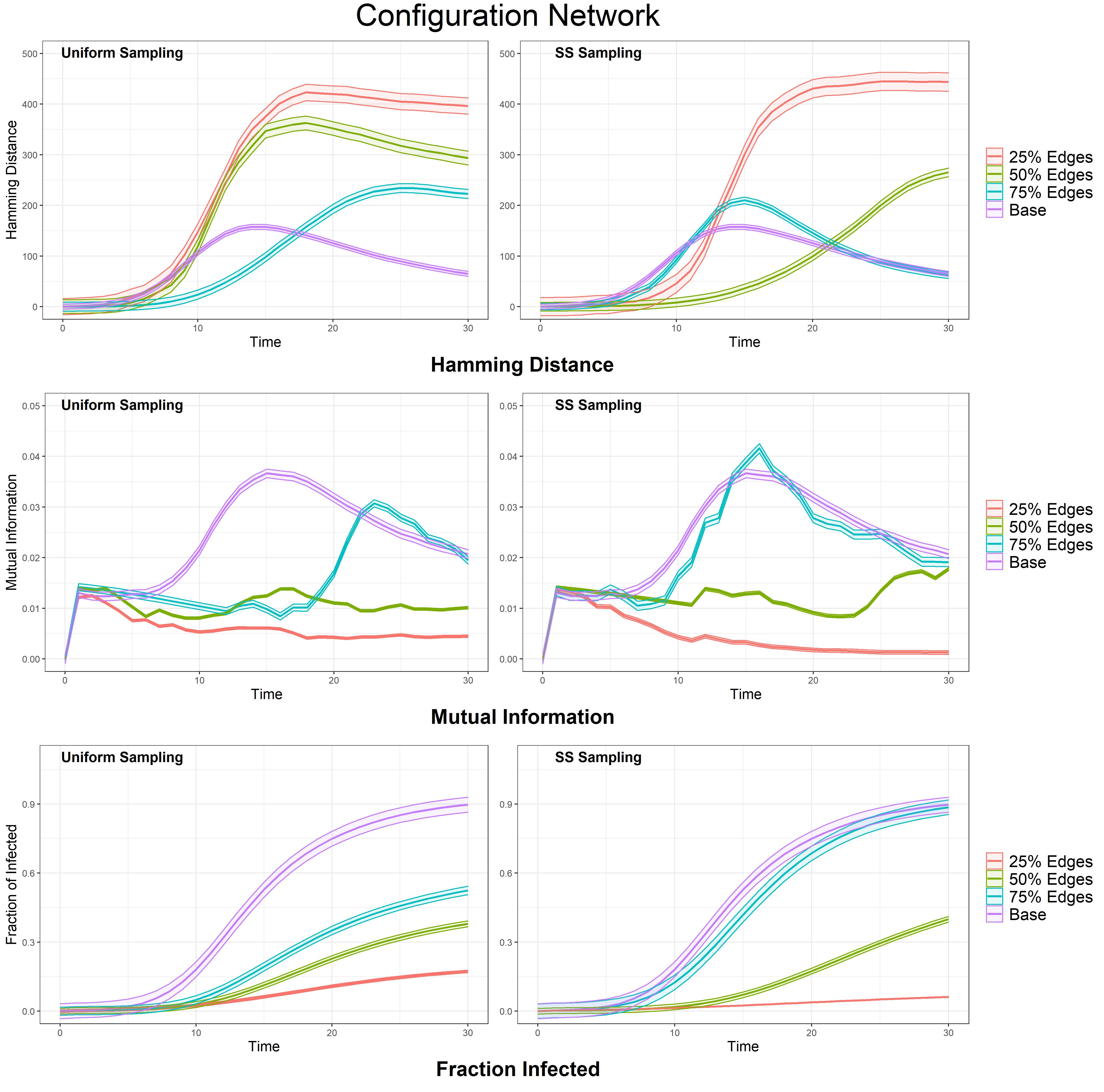}
    \caption{Comparison of CPNS performance on a configuration network with degree list generated from an exponential logarithmic distribution. From top to bottom, the plot displays the Hamming distance, mutual information, and fraction of infected vertices. On the left are the uniform sampling CPNS, termed "Uniform Sampling", and the right the Spielman--Srivastava CPNS, called "SS Sampling." The baseline is shown in purple, $25\%$ edge sparsifier in red, $50\%$ in green, and $75\%$ in blue, with shaded region in each color representing the $95\%$ confidence interval.}
    \label{Figure 1}
\end{figure}

On the configuration network with an exponential-logarithmic degree distribution, the $25\%$ and $50\%$ uniform and Spielman--Srivastava sampling CPNS are comparable across all metrics (Figure\ref{Figure 1}). The $75\%$ Spielman--Srivastava sampling CPNS performs better than the uniform sampling CPNS, adhering closer to the baseline for all metrics (Figure \ref{Figure 1}). It should be noted that the $75\%$ uniform sampling CPNS and $50\%$  Spielman--Srivastava sampling CPNS have lower than baseline Hamming distances (Figure \ref{Figure 1}). This implies that variation found when the SI model was run on the original network was lowered by the CPNS. This could be disadvantageous for the SI model on the CPNS if it wishes to capture the average dynamics found on the original network. 

\newpage

\begin{figure}[H]
    \centering
    \includegraphics[width=\columnwidth]{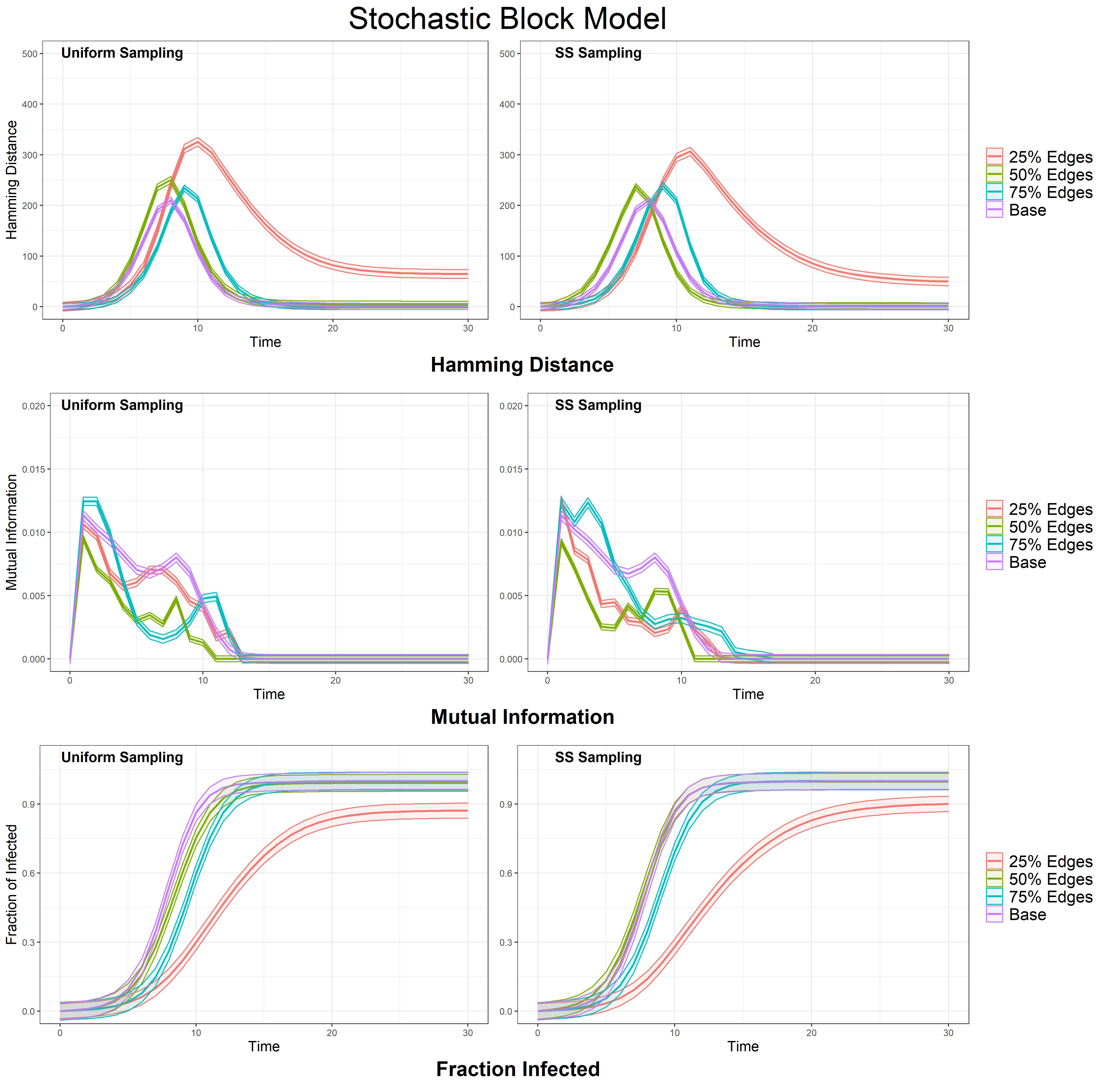}
    \caption{Comparison of CPNS performance on a stochastic block model with four communities of equal size. From top to bottom, the plot displays the Hamming distance, mutual information, and fraction of infected vertices. On the left are the uniform sampling CPNS and the right the Spielman--Srivastava CPNS. The baseline is shown in purple, $25\%$ edge sparsifier in red, $50\%$ in green, and $75\%$ in blue, with shaded region in each color representing the $95\%$ confidence interval.}
    \label{Figure 2}
\end{figure}

For the stochastic block model in Figure 2, the three CPNS of varying edge number for the uniform sampling and SS sampling are comparable. Each of the CPNS adhere closely to the baseline, with the exception of both the uniform sampling and SS sampling $25\%$ CPNS, for both Hamming distance and fraction of infected. No CPNS correctly captured the average mutual information dynamics through time. (Figure \ref{Figure 2}).

\begin{figure}[H]
    \centering
    \includegraphics[width=\columnwidth]{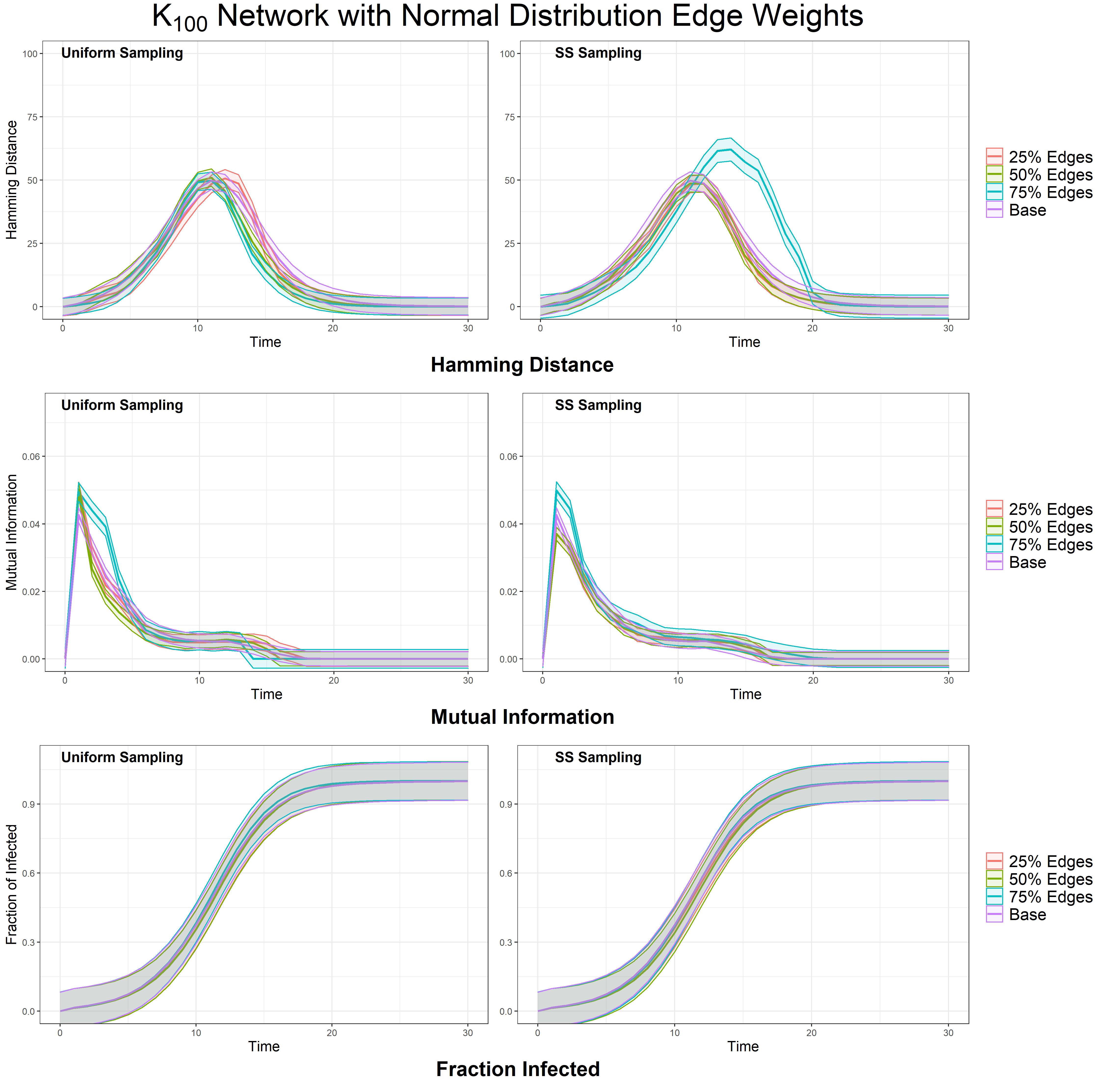}
    \caption{Comparison of CPNS performance on a complete network with $100$ vertices and edge weights drawn from a normal distribution. From top to bottom, the plot displays the Hamming distance, mutual information, and fraction of infected vertices. On the left are the uniform sampling CPNS, termed "Uniform Sampling", and the right the Spielman--Srivastava CPNS, called "SS Sampling." The baseline is shown in purple, $25\%$ edge sparsifier in red, $50\%$ in green, and $75\%$ in blue, with shaded region in each color representing the $95\%$ confidence interval.}
    \label{Figure 3}
\end{figure}

Likewise, for a complete graph with 100 vertices, $K_{100}$, with edge weights drawn from a normal distribution, the three CPNS of varying edge number for the uniform sampling and Spielman--Srivastava sampling are similar, staying close to the baseline. The only exception is the $75\%$ SS sampling CPNS, which has a higher Hamming distance than the baseline in the middle timesteps of the SI simulation (Figure \ref{Figure 3}). However, this elevated average Hamming distance does not effect the $75\%$ Spielman--Srivastava sampling CPNS's performance with either mutual information or the fraction of infected through time. The $75\%$ Spielman--Srivastava sampling CPNS is closer to the mutual information baseline than the corresponding uniform sampling CPNS.

\begin{figure}[H]
    \centering
    \includegraphics[width=\columnwidth]{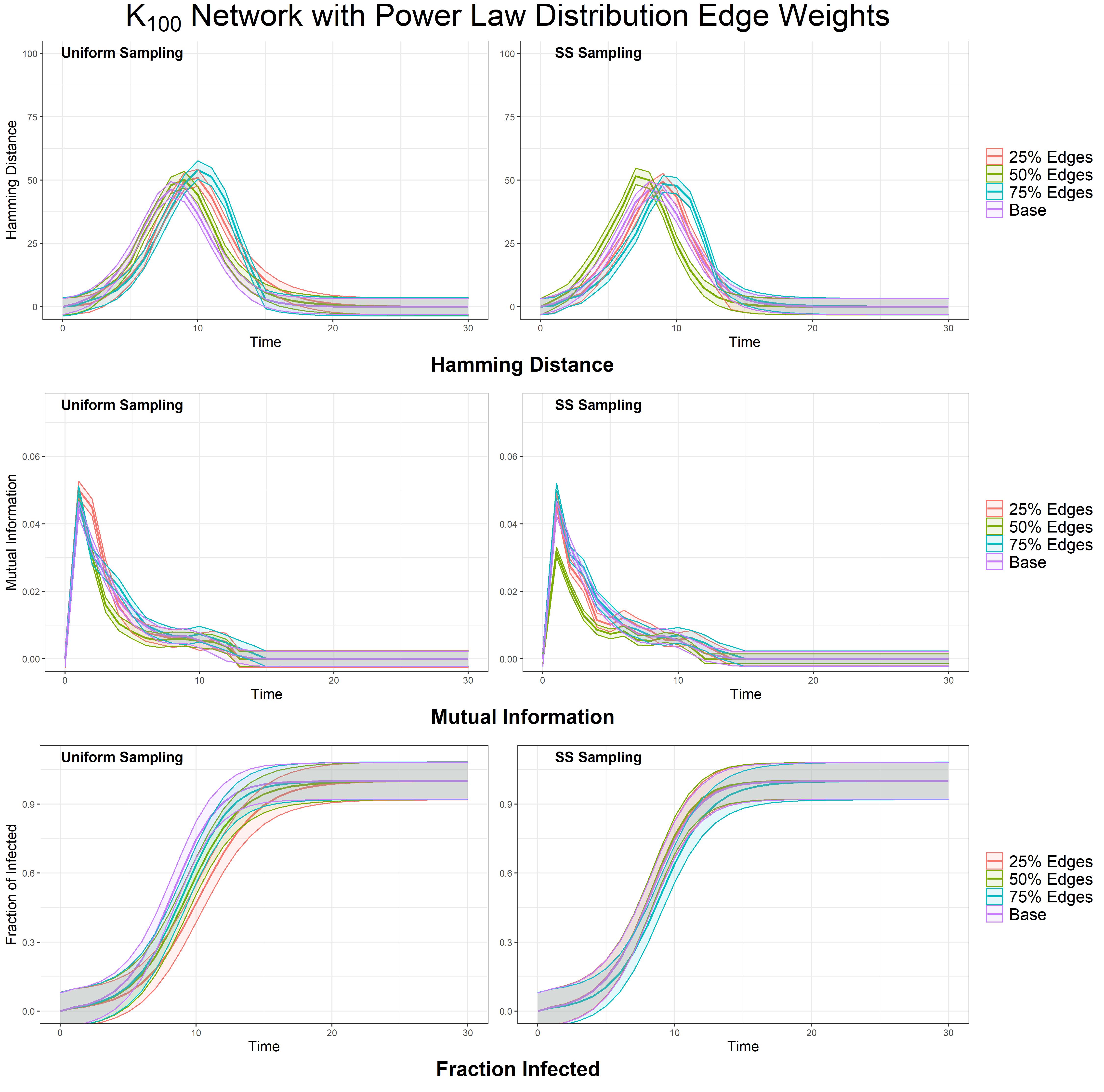}
    \caption{Comparison of CPNS performance on a complete network with $100$ vertices and edge weights drawn from a power law distribution. From top to bottom, the plot displays the Hamming distance, mutual information, and fraction of infected vertices. On the left are the uniform sampling CPNS, termed "Uniform Sampling", and the right the Spielman--Srivastava CPNS, called "SS Sampling." The baseline is shown in purple, $25\%$ edge sparsifier in red, $50\%$ in green, and $75\%$ in blue, with shaded region in each color representing the $95\%$ confidence interval.}
    \label{Figure 4}
\end{figure}

Additionally, the uniform and Spielman--Srivastava sampling CPNS perform similarly on the $K_{100}$ network with edge weights drawn from a power distribution. The $25\%$ and $75\%$ SS sampling are closer to the baseline than the uniform sampling $25\%$ and $75\%$ CPNS (Figure \ref{Figure 4}). However, the $50\%$ Spielman--Srivastava sampling CPNS performs worse with mutual information as a metric than all other CPNS. Lastly, when compared to the uniform sampling CPNS, it appears that the Spielman--Srivastava sampling CPNS is closer to the fraction of infected baseline through time. 

\begin{figure}[H]
    \centering
    \includegraphics[width=\columnwidth]{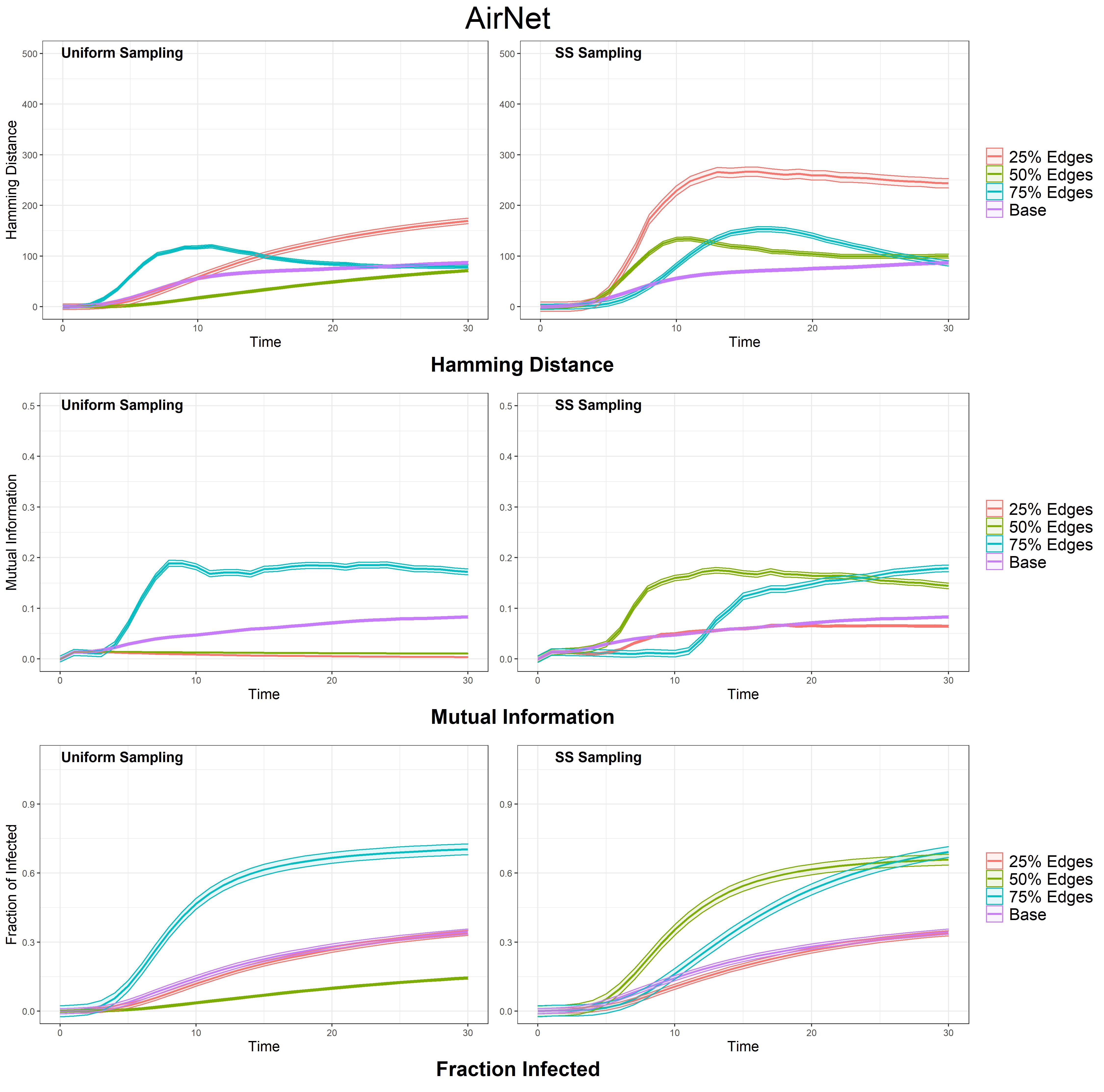}
    \caption{Comparison of CPNS performance on an airline network, AirNet, containing the top $500$ airports in the United States in 2002, with edge weights corresponding to the number of seats passing between pairs of airports. From top to bottom, the plot displays the Hamming distance, mutual information, and fraction of infected vertices. On the left are the uniform sampling CPNS, termed "Uniform Sampling", and the right the Spielman--Srivastava CPNS, called "SS Sampling." The baseline is shown in purple, $25\%$ edge sparsifier in red, $50\%$ in green, and $75\%$ in blue, with shaded region in each color representing the $95\%$ confidence interval.}
    \label{Figure 5}
\end{figure}

Lastly, the we examine a real-world airline network, AirNet. It is notable that neither the uniform nor Spielman--Srivastava sampling CPNS fully capture the SI dynamics on AirNet \ref{Figure 5}. As measured by Hamming distance, it appears that uniform sampling produced more effective CPNS than the Spielman--Srivastava sampling CPNS. Additionally, both $50\%$ uniform and Spielman--Srivastava CPNS and the $75\%$ Spielman--Srivastava CPNS have a larger than baseline fraction of infected through time. Only both $25\%$ CPNS correctly captured the fraction of infected through time. However, as both $25\%$ CPNS performed poorly in Hamming distance and mutual information, regardless of adherence to the baseline, both the uniform and Spielman-Srivastava sampling $25\%$ CPNS fail to capture the totality of dynamics from AirNet. 

\subsection{Comparing Effective Resistance and Epidemic Edge Importance}

\begin{figure}[h]
    \centering
    \includegraphics[width=0.85\columnwidth]{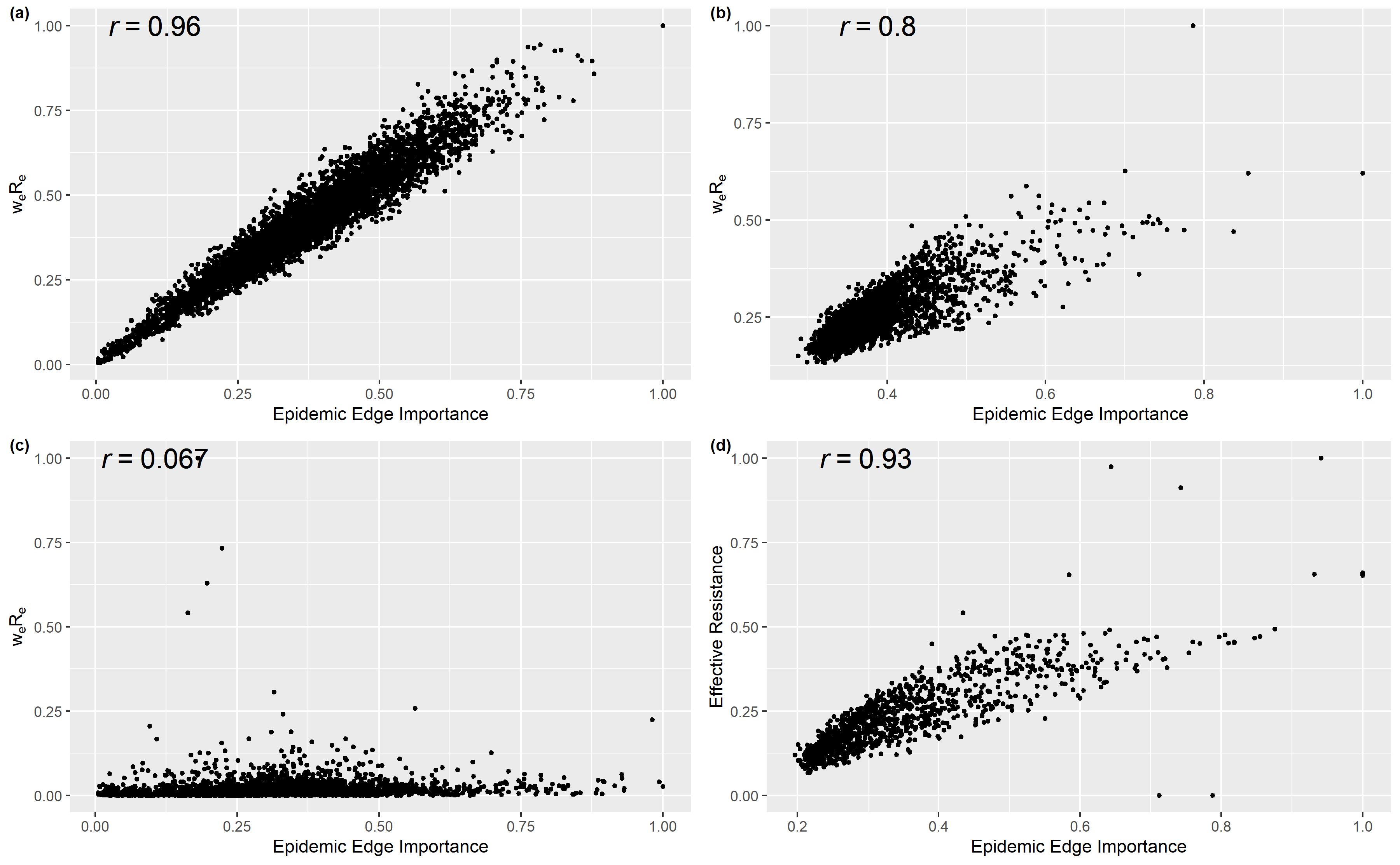}
    \caption{Displayed are the Q-Q plots of the (a) complete network with edge weights from a normal distribution, (b) stochastic block model, (c) AirNet, and (d) configuration network with degree drawn from an exponential-logarithmic distribution. The x-axis corresponds to a normalized epidemic edge importance of edge $e$ and the y-axis to a normalized $w_eR_e$. The value $r$ represents the Pearson correlation coefficient. }
    \label{Figure 6}
\end{figure}

The Q-Q plots showing the similarity of epidemic edge importance and $w_eR_e$ show relatively good correlation for the configuration network $(r=0.93)$, the stochastic block model $(0.8)$, and $K_{100}$ with edge weights selected from a normal distribution $(r=0.96)$  (Figure \ref{Figure 6}). However, AirNet has poor correlation, with $r=0.067$  (Figure \ref{Figure 6}). For AirNet, $w_eR_e$ undervalues the majority of edges deemed important by epidemic edge importance and overvalues certain select edges (Figure \ref{Figure 6}). When considering AirNet, this discrepancy between $w_e R_e$ and epidemic edge importance  seems to be because the majority of highly important $w_eR_e$  edges do not coincide with the edges epidemic edge importance deems as important (Figure \ref{Figure 6}). 

\begin{figure}[H]%
    \centering
    \subfloat[\centering $\gamma=3\times10^{-5}$]{{\includegraphics[width=0.46\columnwidth]{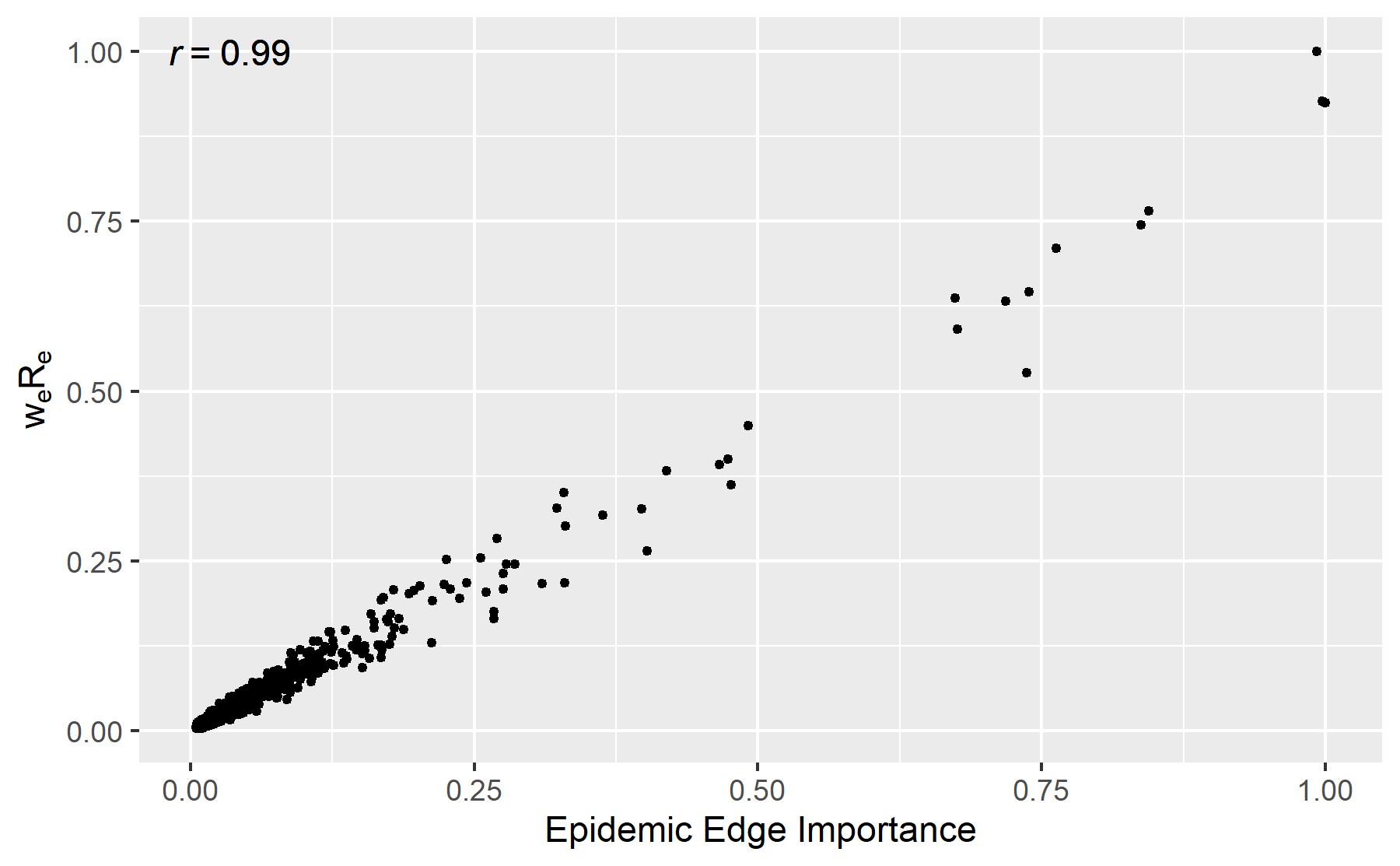} }}%
    \qquad
    \subfloat[\centering $\gamma=3\times10^{-3}$]{{\includegraphics[width=0.46\columnwidth]{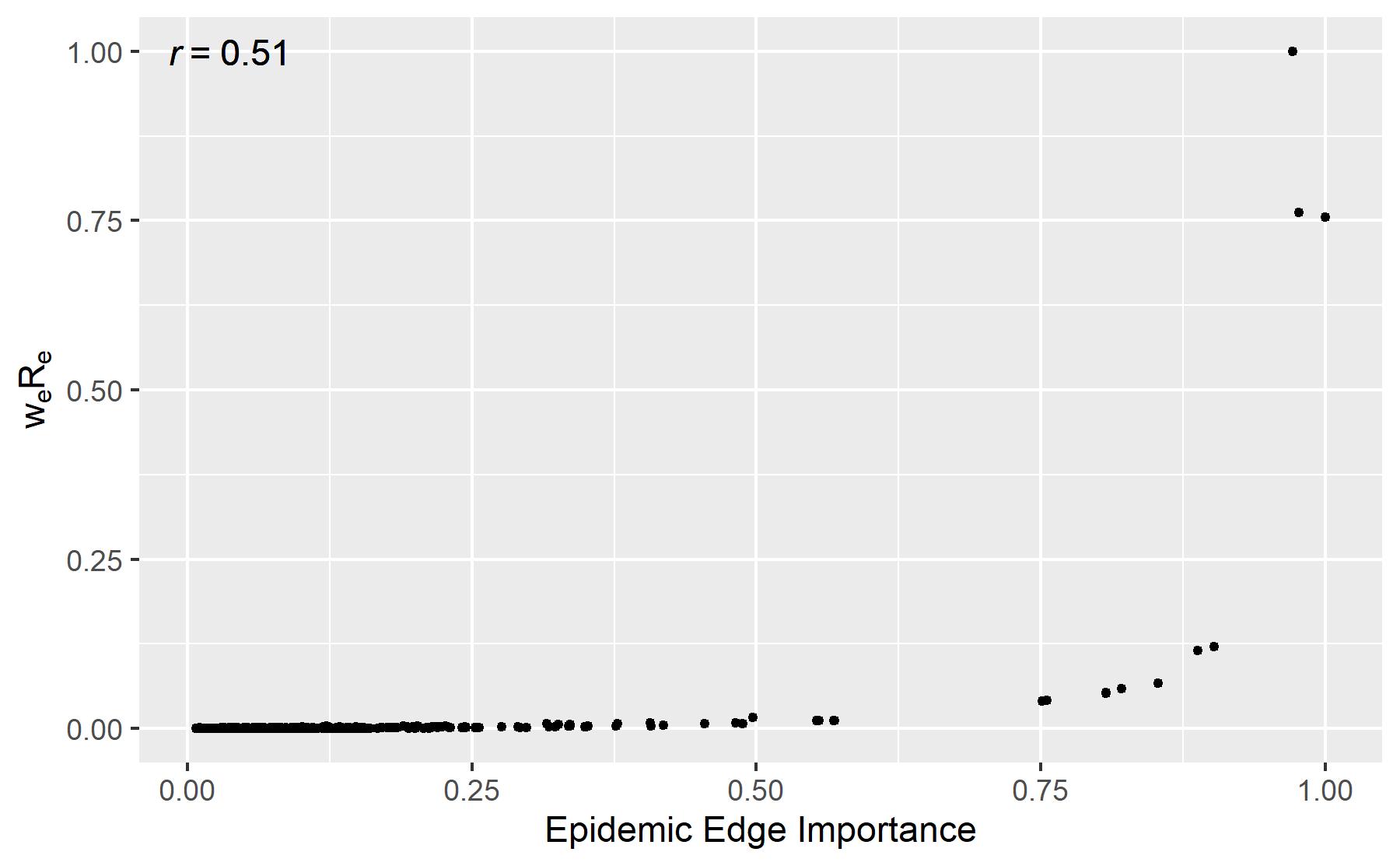} }}%
    \caption{Q-Q Plot of epidemic edge importance and $w_e R_e$ at two different probabilities of transmission, $\gamma$, resulting in two Pearson correlation coefficient values: (a) $r=0.99$ and (b) $r=0.51$.}%
    \label{Figure 7}%
\end{figure}

\begin{figure}[H]%
    \centering
    \subfloat[\centering AirNet: Epidemic Edge Importance]{{\includegraphics[width=0.7\columnwidth]{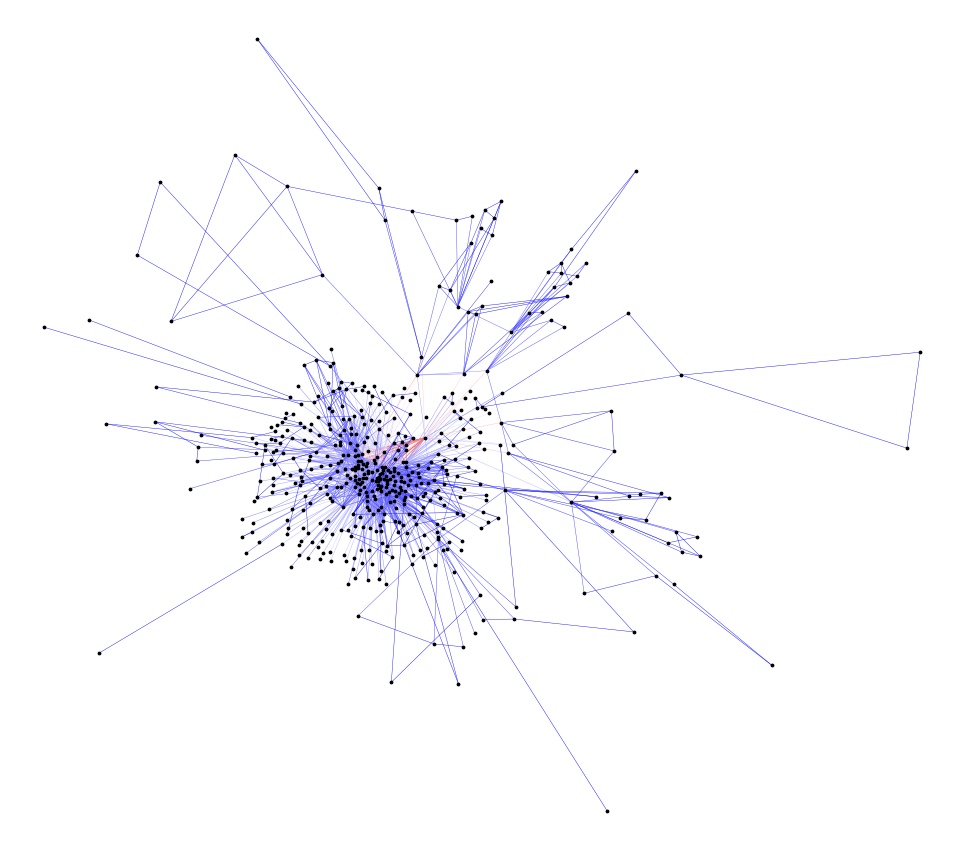} }}%
    \qquad
    \subfloat[\centering AirNet: $w_eR_e$]{{\includegraphics[width=0.7\columnwidth]{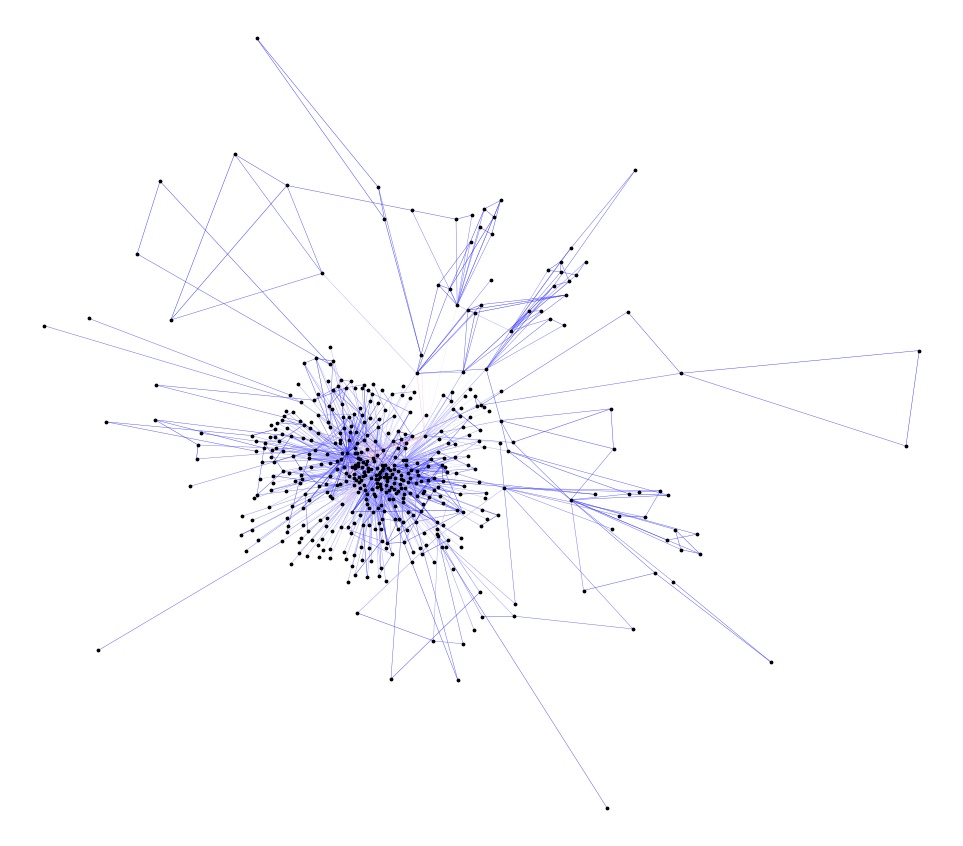} }}%
    \caption{A visualization of the air traffic network AirNet showing (a) epidemic edge importance and (b) $w_eR_e$ . Edges are colored such that red means an edge is more important and blue means an edge is less important for each respective metric.}%
    \label{Figure 8}%
\end{figure}

To further investigate the relationship between epidemic edge importance and $w_eR_e$, two visualizations of AirNet were generated using the NetworkX in Python \cite{hagberg2008exploring}: one with edge color dependent on $w_eR_e$ and another with edge color dependent on epidemic edge importance (Figure \ref{Figure 8}) If an edge has high metric importance, the edge will be colored red. The generation of the two network visualizations suggests a key difference: a subset of edges connecting the core to a singular vertex, which is connected to the periphery of the network, are marked important by epidemic edge importance while $w_e R_e$ does not mark the same subset of edges as important. Instead, the $w_e R_e$ measure of importance  more evenly picks edges throughout the network, with only a few edges in the core of the network being marked as especially important.

Lastly, on the $K_{100}$ network with edge weights drawn from a power law distribution, epidemic edge importance and $w_e R_e$ are poorly correlated ($r=0.51$) for a larger probability of transmission, $\gamma = 3\times 10^{-3}$ , and well correlated ($r=0.99$) for a probability of transmission that is sufficiently small, $\gamma = 3\times 10^{-5}$ (Figure \ref{Figure 7}). Lowering $\gamma$  has two consequences. First, lowering $\gamma$ lowers the possibility that a vertex can be infected simultaneously by two or more of its edges within our SI model. Second, the SI discrete-time model moves closer to a continuous-time model as $\gamma$ is lowered. This suggests that a continuous-time SI model would potentially produce  better correlation between $w_eR_e$ and epidemic edge importance. 

\section{Discussion}

\subsection{Contagion-Preserving Network Sparsifiers and the Spielman-Srivastava Sparsification Algorithm}

To some extent, the Spielman--Srivastava sparsification algorithm successfully created effective CPNS to preserve average SI dynamics across the three metrics with all four random networks. With respect to the configuration network, the Spielman--Srivastava sampling $75\%$ CPNS best adheres to the baseline, allowing for a removal of $25\%$ of the original edges while maintaining approximately the same average SI dynamics as measured by the average Hamming distance, mutual information, and fraction of infected. One quality worth noting is that both the $75\%$ uniform sampling CPNS and the $50\%$ Spielman--Srivastava sampling CPNS have a smaller Hamming distance than the baseline, suggesting that both CPNS lowered the baseline amount of variance between SI runs on the original network and itself. However, by removing some level of variance, this may cause both CPNS to be less faithful to the original network in a probabilistic sense: generating similar distributions of trajectories. 

For the stochastic block model, $50\%$ of the edges could be removed with both the uniform and Spielman--Srivastava sampling, with both $50\%$ CPNS staying close to the baseline in each metric except mutual information where it was lower for both sampling procedures. Both $K_{100}$ networks see effective uniform and Spielman--Srivastava sampling, with the $25\%$ CPNS performing adequately. This allows the removal of $75\%$ of edges in both networks while retaining average SI dynamics. The two aspects of note are the relatively high Hamming distance for the $75\%$ Spielman--Srivastava sampling CPNS on $K_{100}$ with edge weights from a normal distribution and the SS sampling procedure performing marginally better on $K_{100}$ with edge weights from a power law distribution when viewed through the fraction of infected metric. However, while moderately successful CPNS were produced for the four random networks, the Spielman--Srivastava algorithm only unambiguously performed better than the uniform sampling on the configuration network. Moreover, the small size of the networks used in this research were limited by the need to run the SI model over multiple runs on the original network.

This does not necessarily imply that the Spielman--Srivastava performed poorly at generating CPNS. Rather, we suggest that this in part could be explained as a byproduct of the respective network structures. The stochastic block model and complete networks are both well connected, while the configuration network contains more vertices with lower degree. Because this uniform sampling is similar to the performance of the Spielman--Srivastava sampling CPNS such that both are successful at preserving average SI dynamics, this instead implies that edges within those networks have nearly the same level of importance to the epidemic. Conversely, this could be seen that no specific edge is important to the epidemic. In other words, it does not matter which edges are chosen to create the CPNS for those networks. Rather, what matters is that edges \emph{are} chosen for those specific networks. Nevertheless, the relative success of the Spielman--Srivastava sampling procedure on the configuration network when compared to the uniform sampling procedure suggests that there are some instances where the Spielman--Srivastava algorithm will succeed and the uniform sampling procedure will fall short.

The airline network AirNet was the only network where all CPNS that were ineffective. This may be due to how the Spielman--Srivastava algorithm modifies edge weights to compensate for reduced edge number. The Spielman-Srivastava CPNS may be ensuring certain vertices usually infected on the orignal network are almost always infected on the CPNS, inflating the mutual information score above the baseline by removing variation innate to the SI model on the original network. Similarly, the modified edge weights may account for the $50\%$ uniform and Spielman--Srivastava CPNS and the $75\%$ Spielman--Srivastava CPNS having a higher than baseline fraction of infected, where the Spielman--Srivastava algorithm giving higher edge weights to certain edges on the CPNS than found on the orginal network \cite{Swarup:1}. In this way, the Spielman--Srivastava algorithm may cause the contagion to spread faster, causing a higher than baseline fraction of infected and causing the CPNS Hamming distance to not adhere to the baseline.

The relative success of both uniform sampling and Spielman--Srivastava sampling procedures speaks to the effectiveness of random sampling in preserving certain topological features of a network that a deterministic algorithm would not \cite{Swarup:1}. Consider the example of a network with groups that contain many intra-group edges and few inter-group edges such that edges between groups have lower weight than edges within groups. In this scenario, one common deterministic strategy would be to simply remove edges below a certain weight threshold; this would cut off the communities from one another, resulting in a poor CPNS. In contrast, a random sampling (either the uniform sampling or the Spielman--Srivastava sampling procedures) procedure would most likely retain a few of the inter-group edges and produce a better performing CPNS. 

\subsection{Epidemic Edge Importance and Probability of Selection}

The relatively high correlation of $w_eR_e$ with epidemic edge importance -- probability of that same edge appearing in an infection spanning tree -- on the unweighted configuration and stochastic block model networks suggests that the Spielman--Srivastava algorithm is selecting edges with high importance to the SI model. This is especially notable in the configuration network, which is sparser. Similarly, $K_{100}$ with edge weights from a normal distribution also has high correlation between epidemic edge importance and $w_eR_e$. Particularly, for the configuration network, edges connecting low degree vertices have epidemic edge importance and probability of selection close to $1$. Yet, AirNet has relatively low correlation between epidemic edge importance and probability of selection. For AirNet, probability of selection undervalues many edges with high epidemic edge importance, suggesting that the Spielman--Srivastava algorithm is overvaluing certain edges that are not as important to disease spread. Additionally, the correlation between $w_eR_e$ and epidemic edge importance may be dependent on the probability of transmission $\gamma$, whereby if $\gamma$ is small enough the difference in edge weight becomes more pronounced and those bottle necks marked important by effective resistance also have higher epidemic edge importance, as supported by Figure 7.

Because the epidemic edge importance relies on the SI model, the metric of edge importance only relies on the infection rate and the topological structure of the network. This is in contrast to something like the SIR model, where it would be dependent on both the infection and recovery rates. One  consequence of this is that for vertices on the periphery of the network, the SI epidemic will eventually infect them where an SIR metric of edge importance may or may not. This is important when considering all possible patient zeros; An SIR model measure of epidemic edge importance may bias towards the core of the network undervaluing potentially important edges on the periphery. This is ideal when considering potential intervention strategies that necessitate interdiction of an edge. Note that any edge which exists as the only path from one part of the network to another is assigned epidemic edge importance $1$ by our method, as well as a effective resistance of $1$. If this edge leads to only a single isolated vertex on the periphery of the network, this may seem counter intuitive, since a typical epidemic might not reach this vertex. However, in the SI model, all vertices in the connected component containing patient zero eventually becomes infected. Moreover, this isolated vertex might itself be patient zero, in which case its single edge is crucial. 

In general, the idea of epidemic edge importance depends on the details of the epidemic model and parameters used; the probability that any given edge transmit a contagion depends on the specifics of the contagion. For instance,  "SIR epidemic edge importance" could also be defined. We chose not to examine SIR epidemic edge importance, as this measure of edge importance depends on two parameters, rate of infection \emph{and} recovery, instead of one, rate of infection. We focus on the SI model version of epidemic edge importance because it is a simple measure of whether the contagion is likely to spread by an edge if the epidemic reaches (or begins at) either of its endpoints. In this way, the SI version of epidemic edge importance is robust. 

Even so, AirNet had poor correlation between $w_eR_e$ and epidemic edge importance. This may be because even if there are many alternate paths out of the core of the network, those paths may be long or consist of low-weight edges. Then, especially in a discrete-time model where $\gamma$ is fairly large, the epidemic will typically cross to the other part of the network before it has time to traverse the alternate paths. This suggests a tension between vertex centrality and effective resistance of an edge in this particular network, where an edge that is connected to a highly central vertex has high epidemic edge importance but low effective resistance. We see this in the subset of edges with high epidemic edge importance  connecting a vertex which links the core of the network with the periphery of the network. The fact that AirNet is the only network to have poor correlation of $w_eR_e$ and epidemic edge importance and is the only network to produce poor forming CPNS implies that correlation of $w_eR_e$ may be a good predictor of CPNS effectiveness.

\subsection{Future Outlook}

The Spielman--Srivastava algorithm was shown to be reasonably effective at producing reliable and robust contagion-preserving network sparsifiers on an assortment of different networks, allowing for the removal of up to $75\%$ of the edges in certain networks. To some extent, this means that the Spielman--Srivastava algorithm, which approximately preserves the graph Laplacian and therefore linear dynamics, can also approximately preserve the nonlinear behavior of a contagion. However, the fact that even a simple uniform sampling procedure also works fairly well reveals that in some networks there is not a strong enough difference between the importance of different edges to illustrate the specific virtues or faults of the Speilman--Srivastava algorithm within the context of preserving average epidemic dynamics. Moreover, both the uniform and Spielman--Srivastava sampling procedures failed to generate effective contagion-preserving network sparsifiers for AirNet, demonstrating potential flaws. Therefore, a greater variety of networks tailored to illustrate the workings of the Spielman--Srivastava sampling procedure in the context of an epidemic should be considered. 

Additionally, for a majority of the networks considered we found a strong correlation between SI epidemic epidemic importance and the importance, $w_eR_e$, assigned by the Spielman--Srivastava algorithm, suggesting that effective resistance operates as a good approximation of the importance of any given edge in contagion spread. For AirNet, we found that the network that only produced ineffective Spielman-Srivastava CPNS also was the only network that had poorly correlated $w_eR_e$ and epidemic edge importance, suggesting that the correlation between $w_eR_e$ and epidemic edge importance may act as an indicator of Spielman-Srivastava CPNS performance. We also found that this correlation increases when the parameter $\gamma$ in a discrete-time simulation is reduced, or equivalently when we approach a continuous-time version of the SI model, for $K_{100}$ where the edge weights were drawn from a power law distribution. Still, the fact that the real-world airline network did not have well correlated $w_eR_e$ and epidemic edge importance shows that a greater understanding of the relationship between the two metrics is needed. To better explore the notion  of epidemic edge importance, a transition from a discrete time to a continuous time SI model is suggested. More comprehensively, further work is needed to understand the importance of an edge within the context of contagion spread. Intervention strategies acting on vertices, such as vaccinations to protect a vertex, and behavioral interventions, through the interdiction of edges, are critical to  controlling and containing disease spread \cite{Wesolowski:1}.

Much future work remains to address the problem of epidemic sparsifiers. A deeper exploration of network sparsifiers which approximately preserve the simplest case of SI dynamics on larger and more complex real-world networks is needed, as well as an exploration of SIR and more complex epidemic models such as SEIR.  Additionally, while there are appealing parallels between important edges in epidemics and edges with high effective resistance, a better notion of the importance of an edge in the context of an epidemic may exist. Likewise, other notions of edge importance using more complex models than SI should be explored. In general, a deeper understanding of the relationship between linear flows and epidemic processes on networks would greatly benefit this course of research. The question of what edge importance means in the context of contagion spread is critical to producing effective contagion-preserving network sparsifiers . Lastly, a broader testing of other sparsification algorithms within the same class of random sampling sparsifiers to approximately preserve average contagion dynamics is recommended. This leaves the issue of producing effective contagion-preserving network sparsifiers  an open question. 

\section*{Acknowledgments}
\noindent This work was carried out as part of an REU (Research Experience for Undergraduates) program at the Santa Fe Institute under the mentorship of Cristopher Moore and Maria Riolo, funded by NSF grants OAC-1757923 and IIS-1838251. We are also grateful to Samuel Scarpino,  Sayandeb Basu, and Andrew Kramer for their helpful conversations.

\bibliography{CPNS_Paper}
\bibliographystyle{plain}

\end{document}